\documentstyle[aps,12pt]{revtex}
\newcommand{\bb}{\begin{equation}}
\newcommand{\ee}{\end{equation}}
\newcommand{\ba}{\begin{array}}
\newcommand{\ea}{\end{array}}

\begin {document}
\baselineskip 2.2pc

\title{Vacuum polarization for neutral particles 
 in 2+1 dimensions\thanks{published in J. Phys. G {\bf 26} (2000)
 L17-L21.}}
\author{Qiong-gui Lin\thanks{E-mail addresses: qg\_lin@163.net,
stdp@zsu.edu.cn}}
\address{China Center of Advanced Science and Technology (World
	Laboratory),\\
        P.O.Box 8730, Beijing 100080, People's Republic of China\\
        and\\
        Department of Physics, Zhongshan University, Guangzhou
        510275,\\
        People's  Republic of China \thanks{Mailing address}}

\maketitle
\vfill

\begin{abstract}
{\normalsize
In 2+1 dimensions there exists a duality between a charged Dirac
particle coupled minimally to a background vector potential and a
neutral one coupled nonminimally to a background
electromagnetic field strength. A constant uniform background electric
current induces in the vacuum of the neutral particle a fermion
current which is proportional to the background one. A background
electromagnetic plane wave induces no current in the vacuum.
For constant but nonuniform background electric charge,
known results for charged particles can be translated to give the
induced fermion number. Some new examples with infinite background
electric charge are presented. The induced spin and total angular
momentum are also discussed.}
\end{abstract}
\vfill
{\flushleft
PACS number(s): 11.10.Kk, 03.70.+k, 11.15.Tk}

\newpage
\baselineskip 15pt

Vacuum polarization for charged Dirac particles in 2+1 dimensions
has been studied by many authors in recent years [1-14]. The subject
is relevant to many other physical problems in field theory, such as
parity violation, Chern--Simons theory [15-16], flavor symmetry
breaking, etc. The purpose of this note is to extend the subject to
neutral particles. A neutral  particle of spin $\frac 12$
with magnetic moment can interact
with electromagnetic fields through nonminimal coupling.
The Dirac equation for the neutral fermion field $\psi$ is [17]
\bb
(i\gamma^\mu\partial_\mu-\textstyle\frac12\mu_{\rm n}\sigma^{\mu\nu}
F_{\mu\nu}-m_{\rm n})\psi=0,
\ee            
where $m_{\rm n}$ is the mass of the fermion and $\mu_{\rm n}$ its
magnetic moment, $F_{\mu\nu}$ is the external electromagnetic field
strength, $\gamma^\mu$ the Dirac matrices and
\bb
\sigma^{\mu\nu}={i\over 2}[\gamma^\mu, \gamma^\nu].
\ee               
A well known indication of the above nonminimal interaction is the
Aharonov--Casher effect [18] which
has been observed in experiment [19].
In 2+1 dimensions there are two inequivalent irreducible
representations for the Dirac matrices. The first  one is
\bb
\gamma^0=\sigma^3,\quad \gamma^1=i\sigma^1,\quad \gamma^2=i\sigma^2,
\ee     
where the $\sigma$'s are Pauli matrices. In this and all equivalent
representations we have
\bb
\gamma^\mu\gamma^\nu=g^{\mu\nu}-i\epsilon^{\mu\nu\lambda}
\gamma_\lambda,
\ee     
where $g^{\mu\nu}={\rm diag}(1, -1, -1)$, $\epsilon^{\mu\nu\lambda}$
is totally antisymmetric in its indices and $\epsilon^{012}=1$.
The second one is
\bb
\gamma^0=-\sigma^3,\quad \gamma^1=-i\sigma^1,\quad
\gamma^2=-i\sigma^2.
\ee     
In this and all equivalent representations we have
\bb
\gamma^\mu\gamma^\nu=g^{\mu\nu}+i\epsilon^{\mu\nu\lambda}
\gamma_\lambda.
\ee     
On account of the above relations we have
\bb
\sigma^{\mu\nu}=\pm\epsilon^{\mu\nu\lambda}\gamma_\lambda,
\ee     
where the upper (lower) sign corresponds to the first (second)
representation. (The same rule is applied henceforth.) Let us define
\bb
a^\mu=\textstyle\frac 12 \epsilon^{\mu\lambda\rho}F_{\lambda\rho}.
\ee       
Then in 2+1 dimensions Eq. (1) becomes
\bb
[i\gamma^\mu(\partial_\mu\pm i\mu_{\rm n} a_\mu)-m_{\rm n}]\psi=0.
\ee            
Now we write down the Dirac equation for a charged particle in a
background electromagnetic vector potential $A_\mu^e$:
\bb
[i\gamma^\mu(\partial_\mu+ieA_\mu^e)-m_e]\psi_e=0,
\ee            
where we use a subscript (or superscript) $e$ to indicate a charged
particle. Comparing Eqs. (9) and (10) one
easily realizes that $a_\mu$ plays the same role to the neutral
particle as $A_\mu^e$ to the charged one, and $\pm\mu_{\rm n}$
corresponds to $e$. This duality has been
noticed in the study of the Aharonov--Casher effect.
Consequently, constant uniform fields $F_{\mu\nu}$ have no physical
effect on neutral particles. Because of the above
duality, we call $a_\mu$ the dual vecter potential, and
\bb
f_{\mu\nu}=\partial_\mu a_\nu-\partial_\nu a_\mu
\ee     
the dual field strength. Actually, $a_\mu$ is a pseudovector and
$f_{\mu\nu}$ a pseudotensor. Thus there is an essential difference
between Eq. (9) and Eq. (10) in spite of their similar appearance:
Eq. (10) is invariant under the space reflection $(x^0,x^1,x^2)\to
(x^0,-x^1,x^2)$ if $m_e=0$, while Eq. (9) is not even when
$m_{\rm n}=0$.

For charged particles it has been shown [1] by Schwinger's method
of proper time [20] that a current is induced in the vacuum by a
constant uniform background electromagnetic field $F^e_{\mu\nu}$:
\bb
j_e^\mu\equiv\langle 0|{\textstyle\frac 12}[\bar\psi_e, \gamma^\mu
\psi_e]|0\rangle=\pm{e\over 8\pi}\epsilon^{\mu\lambda\rho}
F^e_{\lambda\rho}.
\ee     
It represents the induced electric current in the vacuum
if multiplied by
$e$. By the duality between Eq. (9) and Eq. (10), and using the above
result, we have for the neutral particle a similar one. If
$f_{\mu\nu}$ is constant and uniform, we have an induced fermion
current in the vacuum of the neutral particle:
\bb
j_{\rm n}^\mu\equiv\langle 0|{\textstyle\frac 12}[\bar\psi, \gamma^\mu
\psi]|0\rangle={\mu_{\rm n}\over 8\pi}\epsilon^{\mu\lambda\rho}
f_{\lambda\rho}.
\ee     
Note that the lower (minus) sign in (12) corresponds to the second
representation of the Dirac matrices, and in this representation the
counterpart of $e$ is $-\mu_{\rm n}$, thus the result (13) is
independent
of the choice of the representation. This is different from the case
for charged particles. Substituting Eqs. (8) and (11) into (13) and
using the (2+1)-dimensional Maxwell equation $\partial_\lambda
F^{\lambda\mu}=J^\mu$ where $J^\mu$ is the source current producing
the field $F_{\mu\nu}$, we obtain
\bb
j_{\rm n}^\mu=-{\mu_{\rm n}\over 4\pi}J^\mu.
\ee     
Therefore the induced fermion
current is proportional to the electric source
one. Some remarks on the result: First, the result holds for
constant and uniform $J^\mu$ (or $f_{\mu\nu}$), but $F_{\mu\nu}$ is
necessarily nonuniform or nonconstant. Second, if $A_\mu^e$ (and thus
$F_{\mu\nu}^e$) is a plane wave, we can show that Eq. (12) holds as
well [21]. Consequently, if $F_{\mu\nu}$ (and thus $a_\mu$) is a plane
wave, Eq. (13) or (14) is valid. However, in this case $J^\mu=0$, thus
a background electromagnetic plane wave induces no current. This is
different from the case for charged particles. Third, for
time-independent but nonuniform $J^0$, Eq. (14) does not hold locally,
but an integrated form does hold if the background total charge is
finite. For large total background charge, the local form holds
approximately. But it does not necessarily hold exactly for all
situations with infinite background charge. These are discussed in the
following. For nonuniform {\bf J}, little is known to us at present.

The zero component of Eq. (12) was widely studied in the literature
[2-6]. Though the local form does not hold for nonuniform magnetic
field, an integrated form does hold provided that the total magnetic
flux is finite. Translated to our case, a corresponding result for the
neutral particle read
\bb
Q_{\rm n}\equiv\int j_{\rm n}^0({\bf x})\; d{\bf x}=-{\mu_{\rm n}
Q\over 4\pi},
\ee     
where $Q=\int J^0({\bf x})\; d{\bf x}$ is the total background
electric charge (divided by $e$). That is, the induced fermion
number in the vacuum is proportional to the background electric
charge. For charged particles, of particular interest is the special
case of an Aharonov--Bohm flux string [8-11, 13]. This corresponds
in the neutral case to a pointlike background charge. Eq. (15) holds
in this case according to the previous conclusion for the
Aharonov--Bohm flux string. It is argued that the zero component of
(12) holds approximately as a local relation for large total
magnetic flux [14]. (Though the main attention in Ref. [14] is paid
to the flavor condensate, the argument is applicable to the induced
charge.) Correspondingly, a lacal form of (15) is expected to hold
approximately for large $Q$. However, we will see below that even for
infinite $Q$, the local form does not necessarily hold exactly.

Vacuum induced spin ($S$) and angular momentum ($J$) were also
studied for charged particles in the literature [5-7, 13]. Translated
to our case, we have for finite $Q$ the following results.
\bb
\lim_{m_{\rm n}\to 0} S=\mp{|\mu_{\rm n} Q|\over 8\pi},
\ee     
\bb
J=\mp{(\mu_{\rm n} Q)^2\over 16\pi^2},
\ee     
where the density of spin ($s$) and angular momentum ($j$) is defined
as
\bb
s=\langle 0|{\textstyle\frac 14}[\psi^\dagger, \sigma^3
\psi]|0\rangle,
\ee     
\bb
j=\langle 0|{\textstyle\frac 12}[\psi^\dagger, (l+
{\textstyle\frac 12}\sigma^3)\psi]|0\rangle,
\ee     
where $l=xp_y-yp_x$ is the orbital angular momentum.
Antisymmetrization in the above definition is necessary so that the
expressions are invariant under charge conjugation ($\mu_{\rm n}\to
-\mu_{\rm n}$ for neutral particles).

In the following we consider some special examples with infinite
background electric charge. We denote $\rho=J^0$ and
$\rho_{\rm n}=j_{\rm n}^0$,
and confine ourselves to situations where $\rho$ depends only on a
polar coordinate $r$ such that the Dirac equation can be solved in the
polar coordinates $(r,\theta)$. If $\rho(r)$ behaves like
$r^{-2+\delta_1}$ when $r\to 0$ and like $r^{-2+\delta_2}$ when
$r\to \infty$ with $\delta_1>0$, $\delta_2>0$, then the electric
charge in any finite area is finite but the total charge is divergent.
In these cases there are infinitely many (denumerable) bound-state
solutions with threshold energy $E=m_{\rm n}$ or $-m_{\rm n}$, and no
scattering
solution with the above threshold energies. This is rather different
from those cases with finite background charge, where the number of
threshold bound states is finite and threshold scattering solutions
are also present [22-23]. The current situation with infinite charge
seems simpler. Vacuum polarizations (induced charge, spin, angular
momentum, etc.) are determined by the threshold solutions. Since the
method of calculation is well established in the literature, we only
give the following results.
\bb
\rho_{\rm n}(r)=-{\textstyle\frac 12}{\rm sign}(\mu_{\rm n} q_\infty)
\sum_{m=0}^\infty u_m^2(r),
\ee     
\bb
\lim_{m_{\rm n}\to 0}s(r)=\mp{\textstyle\frac 14}\sum_{m=0}^\infty
u_m^2(r),
\ee     
\bb
j(r)=\mp{\textstyle\frac 12}\sum_{m=0}^\infty(m+{\textstyle\frac 12})
u_m^2(r),
\ee     
where
\bb
u_m(r)=A_m r^m \exp\left[-\left|\int_0^r{\mu_{\rm n} q(r')\over
2\pi r'}\; dr'\right|\right],
\ee     
\bb
q(r)=\int_0^r 2\pi r'\rho(r')\;dr',
\ee     
and $q_\infty$ represents $q(r)$ at large $r$. Obviously, $q(r)$ is
the background charge inside the circle with radius $r$. If $\rho(r)$
has the same sign for all $r$, then ${\rm sign}(\mu_{\rm n} q_\infty)=
{\rm sign}(\mu_{\rm n} \rho)$. The coefficient $A_m$ in Eq. (23) is
determined by the normalization condition
\bb
\int_0^\infty 2\pi r u_m^2(r)\; dr=1.
\ee     
From the above results we see that there exists a local relation
between $\rho_{\rm n}(r)$ and $\lim_{m_{\rm n}\to 0}s(r)$:
\bb
\lim_{m_{\rm n}\to 0}s(r)=\pm{\textstyle\frac 12}{\rm sign}
(\mu_{\rm n} q_\infty)\rho_{\rm n}(r).
\ee     
It is not clear whether this relation holds in other charge
configurations without cylindrical symmetry. It is also not clear
whether it holds in situations with finite background charge, because
the contribution from scattering threshold states is subtle. However,
an integrated relation does hold when the background charge is finite.
[cf. Eqs. (15) and (16).]

When $\rho(r)\propto r^{-2+\delta}$ with $\delta>0$, exact expressions
are available for all the above quantities. However, closed forms can
be acquired only when $\delta=1,2,4$.
It is worth noting that the Dirac equation can be completely solved
when $\delta=1,2$.
The case $\delta=2$ corresponds
to the one with constant uniform charge density discussed above.
Calculation of $\rho_{\rm n}(r)$ by using the above expressions
confirms the previous result (14) (the zero component).
The results for the other two cases are given below.
Since $\lim_{m_{\rm n}\to 0}s(r)$ is related to $\rho_{\rm n}(r)$ by
(26), and $j(r)$ is not very interesting (and also
complicated when $\delta=4$), we only write down the results for
$\rho_{\rm n}(r)$. When $\rho(r)=\lambda_1/r$, we have
\bb
\rho_{\rm n}(r)=-{\mu_{\rm n}\rho(r)\over 4\pi}[1-
\exp(-4|\mu_{\rm n}\lambda_1|r)],
\ee     
while when $\rho(r)=\lambda_2 r^2$, we have
\bb
\rho_{\rm n}(r)=-{\mu_{\rm n}\rho(r)\over 8\pi}\left[1+{\rm erf}
\left(\sqrt{|\mu_{\rm n}\lambda_2|\over 8}r^2\right)\right]-
{\rm sign}(\mu_{\rm n}\lambda_2)
{\sqrt{|\mu_{\rm n}\lambda_2|}\over (2\pi)^{\frac 32}}\exp\left
(-{|\mu_{\rm n}\lambda_2|\over 8}r^4\right),
\ee     
where $\lambda_1$, $\lambda_2$ are constants, and
$$
{\rm erf}(\xi)={2\over\sqrt\pi}\int_0^\xi e^{-\eta^2}\; d\eta.
$$
Although the background charge is infinite, the local relation (14)
(the zero component) is not valid. It holds approximately only when
$\lambda_1$ or $\lambda_2$ is large. The above results can be
translated to the case of charged particles. However, it should be
pointed out that the significance of these examples is limited since
the situation is difficult to realize in practice.

As in the case of charged particles, one may consider a four-component
form of the theory with $\gamma^0={\rm diag}(\sigma^3,-\sigma^3)$,
$\gamma^1={\rm diag}(i\sigma^1,-i\sigma^1)$,
$\gamma^2={\rm diag}(i\sigma^2,-i\sigma^2)$.
It is then invariant under the space reflection
$(x^0,x^1,x^2)\to(x^0,-x^1,x^2)$. In this case, the induced spin and
angular momentum in the vacuum both vanish. But the induced charge is
twice that given above, which is rather different from the case of
charged particles. Further more, there is no flavor or chiral symmetry
even when $m_{\rm n}=0$, which is also different from the case of
charged particles. Therefore it is meaningless to speak about flavor
symmetry breaking in this theory.

Recently, the duality between neutral and charged particles has been
used to calculate the probability of neutral particle-antiparticle
pair creation in the vacuum
by external electromagnetic fields in 2+1 dimensions [24].
Unfortunately, nontrivial results
in 3+1 dimensions are still not available. On the other hand,
Eq. (1) has been solved in some spherically symmetric electric fields
in 3+1 dimensions [25], and spectral asymmetry
similar to that in 2+1 dimensions is found. Therefore vacuum
polarizations for neutral particles in 3+1 dimensions can be
expected. This will be studied subsequently.

\vskip 1cm

This work was supported by the
National Natural Science Foundation of China.


\end{document}